\documentclass[12pt]{iopart}
\usepackage{epsf}
\usepackage{iopams}
\usepackage[dvips]{graphicx}
\usepackage{rotating}

\newcommand{\bd     }{\begin{displaymath}}
\newcommand{\ed     }{\end{displaymath}}

\newcommand{\s      }{\sigma}

\newcommand{\bra    }{\langle}
\newcommand{\ket    }{\rangle}

\newcommand{\atanh  }{{\rm{ath}}}

\begin{document}

\title
[The Little-Hopfield model on a Random Graph]
{The Little-Hopfield model on a Random Graph}
\author{ I.~P\'{e}rez Castillo and N.S.~Skantzos}
\address{Institute for Theoretical Physics, Celestijnenlaan 200D,
Katholieke Universiteit Leuven, Leuven, B-3001 Belgium}


\begin{abstract}
\noindent
We study the Hopfield model on a random graph in scaling regimes where 
the average number of connections per neuron is a
finite number and where the spin dynamics is governed by a synchronous
 execution of the microscopic update rule (Little-Hopfield model).
We solve this model within replica symmetry and by using bifurcation analysis
we prove that the spin-glass/paramagnetic and the retrieval/paramagnetic 
transition lines
of our phase diagram are identical to those of sequential dynamics. 
The first-order
retrieval/spin-glass transition line follows 
by direct evaluation of our observables
using population dynamics. Within the accuracy of numerical precision 
and for sufficiently small values of the connectivity parameter 
we find that this line coincides with the corresponding sequential one. 
Comparison with simulation experiments shows excellent agreement.
\end{abstract}

\section{Introduction}

Since the first analytical work describing pattern recall
was presented and the theoretical foundations describing the
operation of neural networks were subsequently set,
the progress in the field of attractor neural network models has been
advancing revealing interesting properties. Early analytical attempts
to solve non-trivial neural network systems were analytically constrained
to the study of fully-connected ones. From a real
(i.e.\@ biological) point of view one would ideally prefer to study models 
of sparse connectivity. Subsequently, theoretical work turned (among other directions)
to sparse models, and, in particular, to the so-called
extremely diluted ones \cite{canning}. These could be still solved exactly
owing to the simplification of considering system sizes
exponentially bigger than the average connectivity per neuron (while
both of these numbers were eventually sent to infinity). 

One direction in developing the neural network theory further towards realism
implies moving away from the simplifications made in extremely
diluted systems and considering \emph{finite} degrees of connectivity.
This however appears to be a highly non-trivial step ahead
since at an early stage of solving the relevant equations one
is confronted with more than the traditional two observables
(the magnetisation and the overlap), in fact, one is required
to consider an order-parameter function.

Although initial attempts to solve such models showed the underlying analytical
complexity \cite{viana,kanter}, the more refined mathematical tools were
only relatively recently developed and in fact for
applications different than neural networks (for instance,
for optimisation problems \cite{optim} or error-correcting codes 
\cite{error1}).
The first study which applied the finite-connectivity methodology
to neural network problems has been very recently presented in \cite{bastian}, 
where phase diagrams were presented for a variety of synaptic kernels.
There, the authors study neural networks in which the microscopic
neuronal dynamics is a sequential execution of an update rule
(describing alignment to post-synaptic potentials).
In this paper we examine the effect of choosing a \emph{synchronous}
microscopic dynamics in Hopfield models of finite connectivity,
i.e.\@ one in which neurons are updated
in parallel at each time step. These two different types
of dynamics are known to share very interesting properties.
For instance, in simple ferromagnetic models one can prove
that thermodynamic observables become identical \cite{muller}. 
It is yet unclear to what extent the two types of dynamical models
share these common equilibrium features. 
For example, it is known that the phase 
diagram of the synchronous Hopfield model changes 
\cite{toni,fontanari}\footnote{
This difference seems to be an artifact of the replica symmetric approximation.
The two phase diagrams become identical within full Replica 
Symmetry Breaking \cite{toni}, although it is unclear at which stage of the 
breaking scheme this occurs.}
whereas that of the Sherrington-Kirkpatrick model remains
unaffected \cite{nishimori}. 

This paper is organised as follows: in the following section we present
definitions of our model. In section \ref{sec:saddle} we
derive the saddle point equations for the free energy
and make our replica-symmetric ansatz. In section \ref{sec:selfcons} we
give expressions of the free energy in terms of the replica symmetric
order function whereas our results and phase diagrams are given
in section \ref{sec:results}.

\section{Definitions }
\label{sec:def}

We study neural network models of $N$
binary neurons $\bsigma=(\s_1,\ldots,\s_N)$
with $\s_i=-1$ representing `at rest'  and $\s_i=1$
the `firing' state.
The miscroscopic neuron dynamics
is a stochastic alignment to `local fields' (the post-synaptic potentials)
in which updates in
neuronal states are made for all $i\in\{1,\ldots,N\}$
in a fully synchronous way at each discrete
time step:
\begin{equation}
{\rm Prob}[\s_i(t+1)=\pm 1]
=\frac12[1\pm\tanh(\beta h_i(\bsigma(t)))]
 \label{eq:micro_dynamics}
\end{equation}
where $h_i(\bsigma(t))=\sum_j J_{ij}\s_j(t)$.
The parameter $\beta\in[0,\infty)$ controls the amount of
thermal noise  in the dynamics with $\beta=\infty$ corresponding
to a fully deterministic execution of (\ref{eq:micro_dynamics})
and $\beta=0$ to a fully random execution.
The quantities $J_{ij}$ describe interaction couplings.
If one expresses (\ref{eq:micro_dynamics})
in terms of the
microscopic state probabilities $p_t(\bsigma)$:
\begin{equation}
p_{t+1}(\bsigma)=\sum_{\bsigma'}W[\bsigma;\bsigma']p_t(\bsigma')
\hspace{5mm}
W[\bsigma;\bsigma']=\prod_{i=1}^N
\frac{e^{\beta \s_i h_i(\bsigma')}}
     {2\cosh[\beta \s_i h_i(\bsigma')]}
\label{eq:dynamics}
\end{equation}
then, for any finite $\beta$ and finite $N$ the
process (\ref{eq:dynamics}) is ergodic and evolves to a unique
distribution $p_\infty(\bsigma)$. It can be shown that
this is a unique equilibrium state (obeying detailed balance)
if and only if $J_{ij}=J_{ji}$ \cite{peretto}
\footnote{Note that unlike sequential models where detailed balance
requires the exclusion of self-interactions, this is not here a prerequisite. 
However, due to the scaling regime of finite connectivity terms originating 
from the self-interacting part will for $N\to\infty$ give a
vanishing contribution to thermodynamic quantities.}. At equilibrium, the
microscopic state probabilities acquire the form
$p_\infty\sim \exp[-\beta H(\bsigma)]$, where
\begin{equation}
H(\bsigma)=-\frac1\beta \sum_i\log[2\cosh(\beta h_i(\bsigma))]
\label{eq:H}
\end{equation}
is the Hamiltonian. For the interactions $J_{ij}$ we now take:
\begin{equation}
J_{ij}=\frac{c_{ij}}{c}\sum_{\mu=1}^p
\xi_i^\mu\xi_j^\mu
\label{eq:interactions}
\end{equation}
This corresponds to storing $p$ memories
$\bxi_i=(\xi_i^1,\ldots,\xi_i^p)\in\{-1,1\}^p$
among the synapses in a Hebbian-type way.
The variables $c_{ij}\in\{0,1\}$ with $c_{ij}=c_{ji}$ represent
dilution, while $c=\sum_{j}c_{ij}$ (for all $i$) corresponds to the
average number of connections per neuron.
Models of the type (\ref{eq:interactions})
with  $c\to\infty$ (while $c\sim \log N$) are known as
extremely diluted and due to their simplicity have been studied
extensively (see for instance \cite{bolle,theumann,derrida}
 and references therein). What is less known are properties of these systems
in the non-trivial scaling regime of
\emph{finite connectivity} where $c\sim\mathcal{O}(1)$
(with the probability $c/N\to 0$).
Due to the complicated methodology required
such systems have only recently been studied in \cite{bastian}
where a thorough bifurcation analysis was performed
and phase diagrams were derived.
For the distribution of the random variable $c_{ij}$
we will consider
\begin{equation}
P(c_{ij})=\frac{c}{N}\delta_{c_{ij},1}+\left(1-\frac{c}{N}\right)
\delta_{c_{ij},0}
\label{eq:avc}
\end{equation}
for all pairs $(i,j)$. All thermodynamic quantities
will have to be averaged over (\ref{eq:avc}).
Note that due to the system's sparse connectivity
the number of patterns can only be finite.

To derive observable quantities we will calculate the free energy
per neuron $f=-\lim_{N\to\infty}\frac{1}{\beta
N}\bra\,\log\sum_{\bsigma}e^{-\beta H(\bsigma)}\,\ket_{\{c_{ij}\}}$,
with $\bra \cdots\ket$ denoting average over the distribution of dilution variables.
 As in all synchronous dynamics models, the evaluation of the free energy is greatly
simplified by introducing an extra set of spins so that $f$ can be
rewritten as
\begin{equation}
\fl
f=
-\lim_{N\to \infty}\frac{1}{\beta N}\left\bra\log \sum_{\bsigma\btau}
e^{-\beta \mathcal{H}(\bsigma,\btau)}\right\ket_{\{c_{ij}\}}
\hspace{10mm}
\label{eq:f_def}
\mathcal{H}(\bsigma,\btau)=-\sum_{ij}\s_i J_{ij} \tau_j
\end{equation}
Equation (\ref{eq:f_def}) will be the starting point for our analysis.

\section{Saddle-point equations}
\label{sec:saddle}

In order to calculate the free energy (\ref{eq:f_def}) we
begin by invoking the
replica identity $\bra \log Z\ket=\lim_{n\to 0} \frac1n\log \bra
Z^n\ket$. One can then take the average over the dilution
variables resulting in
\begin{equation}
\fl
f
=
-\lim_{N\to\infty}\lim_{n\to 0}\frac{1}{\beta Nn}\log
\sum_{\bsigma^1\cdots\bsigma^n}
\sum_{\btau^1\cdots
\btau^n}\exp\left[\frac{c}{2N}\sum_{ij}\left(e^{\frac{\beta}{c}
(\bxi_i\cdot\bxi_j)
\sum_{\alpha}(\s^\alpha_i\tau^\alpha_j+
\s^\alpha_j\tau^\alpha_i)}-1\right)\right]
\end{equation}
where  $\bxi_i\cdot\bxi_j=\sum_\mu\xi^\mu_i\xi_j^\mu$.
Note that at this stage an exponential has appeared as an argument
of another exponential. Here, effectively, the inner-most
exponential introduces an infinite number of observables. However, unlike systems of
full-connectivity or of extreme dilution where only the linear and quadratic
 observables (order parameters) {\it survive}  in the limit $N\to\infty$, 
 all cumulants play here a role. They can be re-cast in an order-parameter function.
This complication is a direct consequence of the
scaling regime of finite connectivity. 

It is convenient now to use
the concept of sublattices \cite{vanhemmen} which divide the space
of sites into $2^p$ groups $I_{\bxi}=\{i|\bxi_i=\bxi\}$. We define
$\bsigma_i=(\s_i^1,\ldots,\s_i^n)$ and abbreviate the averages over
the sublattices by $\bra F(\bxi)\ket_{\bxi}=\sum_{\bxi} p_{\bxi}
F(\bxi)$ with the probabilities $p_{\bxi}\equiv |I_{\bxi}|/N$. Then
the free energy becomes:
\[
\fl f=
-\lim_{N,n}\frac{1}{\beta Nn}\log\sum_{\bsigma^1\cdots\bsigma^n}
\sum_{\btau^1\cdots \btau^n}
\exp\left[\frac{c}{2N}\sum_{\bsigma\bsigma'}\sum_{\btau\btau'}
\sum_{\bxi\bxi'}
\left(e^{\frac{\beta }{c}(\bxi\cdot\bxi')\sum_\alpha
(\s_\alpha\tau_\alpha^\prime+\s^\prime_\alpha\tau_\alpha)}-1\right)
\right.
\]
\begin{equation}
\hspace{20mm}
\times\left.
\sum_{i\in I_{\bxi}}\delta_{\bsigma,\bsigma_i}\delta_{\btau,\btau_i}
\sum_{j\in I_{\bxi'}}\delta_{\bsigma',\bsigma_j}\delta_{\btau',\btau_j}\right]
\label{eq:f}
\end{equation}
We now
see that an order-function has emerged which we introduce into our
expression via
\begin{equation}
1=\int\prod_{\bxi\bsigma\btau}\left\{d P_{\bxi}(\bsigma,\btau)\
\delta\left[P_{\bxi}(\bsigma,\btau)-\frac{1}{|I_{\bxi}|}\sum_{i\in I_{\bxi}}
\delta_{\bsigma,\bsigma_i}\delta_{\btau,\btau_i}\right]\right\}
\end{equation}
In the above expression we replace the $\delta$-function by its 
integral representation (which 
introduces the conjugate order function parameter 
$\hat{P}_{\bxi}(\bsigma,\btau)$). We then perform the trace over the spin
variables and take the limit $N\to\infty$ in our equations
resulting in the extremisation problem:
\begin{equation}
\fl 
f=-\lim_{n\to 0}\frac{1}{\beta n}{\rm Extr}_{\{P,\hat{P}\}}
\left\{\bra\sum_{\bsigma\btau}P_{\bxi}(\bsigma,\btau)\hat{P}_{\bxi}
(\bsigma,\btau)\ket_{\bxi}+
\bra\,\log[\sum_{\bsigma\btau}e^{-\hat{P}_{\bxi}(\bsigma,\btau)}]\,
\ket_{\bxi}
\right.
\end{equation}
\begin{equation}
\left.+\frac{c}{2}\bra\sum_{\bsigma\btau}\sum_{\bsigma'\btau'}
P_{\bxi}(\bsigma,\btau)P_{\bxi'}(\bsigma',\btau')
\left[e^{\frac{\beta}{c}(\bxi\cdot\bxi')
\sum_{\alpha}(\s_\alpha\tau'_\alpha+
\s'_\alpha\tau_\alpha)}-1\right]\ket_{\bxi\bxi'}\right\}
\end{equation}
Variation with respect to the densities $P_{\bxi}(\bsigma,\btau)$
and $\hat{P}_{\bxi}(\bsigma,\btau)$ gives the self-consistent
equation
\begin{equation}
\fl 
P_{\bxi}(\bsigma,\btau)=
\frac{\exp\left[c\bra\sum_{\bsigma'\btau'}P_{\bxi'}(\bsigma',\btau')
\left(e^{\frac{\beta}{c}(\bxi\cdot\bxi')\sum_\alpha(\s_\alpha\tau^\prime_\alpha
+\s^\prime_\alpha\tau_\alpha)}-1\right)\ket_{\bxi'}\right]}
{\sum_{\bsigma'\btau'}
\exp\left[c\bra\sum_{\bsigma''\btau''}P_{\bxi''}(\bsigma'',\btau'')
\left(e^{\frac{\beta}{c}(\bxi\cdot\bxi'')
\sum_\alpha(\s^\prime_\alpha\tau^{\prime\prime}_\alpha
+\s^{\prime\prime}_\alpha\tau_\alpha^\prime)}-1\right)\ket_{\bxi''}\right]}
\label{eq:selfcons1}
\end{equation}
and also allows us to eliminate $\hat{P}_{\bxi}(\bsigma,\btau)$ from the
expression of the free energy which now becomes:
\[
\fl
f=\lim_{n\to 0}\frac{1}{\beta n}{\rm Extr}\left\{
\frac{c}{2}\bra\sum_{\bsigma\btau}\sum_{\bsigma'\btau'}
P_{\bxi}(\bsigma,\btau)P_{\bxi'}(\bsigma',\btau')
\left(e^{\frac{\beta }{c}(\bxi\cdot\bxi')
\sum_\alpha(\s_\alpha\tau^\prime_\alpha
+\s^\prime_\alpha\tau_\alpha)}-1\right)\ket_{\bxi\bxi'}\right.
\]
\begin{equation}
\left.
-\left\bra\log\sum_{\bsigma\btau}\exp\left[c\Big\bra\sum_{\bsigma'\btau'}
P_{\bxi'}(\bsigma',\btau')
\left(e^{\frac{\beta }{c}(\bxi\cdot\bxi')
\sum_\alpha(\s_\alpha\tau^{\prime}_\alpha
+\s^{\prime}_\alpha\tau_\alpha)}-1\right)\Big\ket_{\bxi'}\right]\right\ket_{\bxi}\right\}
\label{eq:f2}
\end{equation}
This expression requires knowledge of the densities
$P_{\bxi}(\bsigma,\btau)$.
However, the evaluation of the self-consistent equation (\ref{eq:selfcons1})
is an impossible task unless a further simplification is
made. To proceed further we will make an assumption about the form
of the densities $P_{\bxi}(\bsigma,\btau)$.
In the spirit of replica symmetry (RS) we will consider that
permutation of spins within different replicas leave the order
parameters invariant. Here however, due to the presence
of two species of spins
we will also require that within the same replica group
permutation of the states of $\s_\alpha$ and $\tau_\alpha$
also leave for all $\alpha$ $P_{\bxi}(\bsigma,\btau)$ invariant:
\begin{equation}
P_{\bxi}(\bsigma,\btau)=\int \frac{dhdrdt}{[\mathcal{N}(h,r,t)]^n}\
W_{\bxi}(h,r,t)\
e^{\beta h\sum_\alpha\s_\alpha+\beta r\sum_\alpha \tau_\alpha
+\beta t\sum_\alpha\s_\alpha\tau_\alpha}
\label{eq:RS}
\end{equation}
where $\mathcal{N}(h,r,t)$ is the corresponding normalisation constant
\begin{equation}
\fl
\mathcal{N}(h,r,t)=4\cosh(\beta h)\cosh(\beta r)\cosh(\beta t)
+4\sinh(\beta h)\sinh(\beta r)\sinh(\beta t)
\end{equation}

Let us finally turn to our system's macroscopic observables. Our
replicated sublattice overlaps will be given by
\begin{equation}
m^{\mu\alpha}_{\lambda}=\bra \xi^\mu m_{\bxi}^{\lambda,\alpha}\ket_{\bxi}
\label{eq:mlambda}
\end{equation}
 with the sublattice magnetisations defined as
\begin{equation}
m_{\bxi}^{\lambda,\alpha}=\sum_{\bsigma\btau}P_{\bxi}(\bsigma,\btau)\lambda^\alpha
\end{equation}
with $\lambda=\s,\tau$. In RS the above quantities are the same 
for all values of the replica index. Thereafter we
 will drop for notational simplicity the index $\alpha$.
The observables (\ref{eq:mlambda}) will be generated from 
the densities $P_{\bxi}(\bsigma,\btau)$. In general, from $P_{\bxi}(\bsigma,\btau)$
one can evaluate all higher order observables
\begin{equation}
L_{\bxi}^{\alpha_1\cdots\alpha_m;\gamma_1\cdots\gamma_\ell}=\sum_{\bsigma\btau}
P_{\bxi}(\bsigma,\btau)\ \s^{\alpha_1}\cdots\s^{\alpha_m}
\tau^{\gamma_1}\cdots\tau^{\gamma_\ell}
\end{equation}
Working out the simplest of these using the RS ansatz (\ref{eq:RS}) gives:
\begin{eqnarray}
m_\s^{\mu}
& \equiv &
\bra \xi^\mu \sum_{\bsigma\btau}P_{\bxi}(\bsigma,\btau)\s^{\alpha_1}\ket_{\bxi}
\nonumber
\\
&= &
\bra \xi^\mu \int dhdrdt\,
W_{\bxi}(h,r,t)\ \frac{\tanh(\beta h)+\tanh(\beta r)\tanh(\beta t)}
{1+\tanh(\beta t)\tanh(\beta r)\tanh(\beta h)}\ket_{\bxi}
\label{eq:ms}
\end{eqnarray}
\begin{eqnarray}
q_{\s\s}
&\equiv &
\bra \sum_{\bsigma\btau}P_{\bxi}(\bsigma,\btau)\ \s^{\alpha_1}\s^{\alpha_2}
\ket_{\bxi}
\nonumber
\\
& = &
\bra \int dhdrdt\,
W_{\bxi}(h,r,t)\ \left[\frac{\tanh(\beta h)+\tanh(\beta r)\tanh(\beta t)}
{1+\tanh(\beta t)\tanh(\beta r)\tanh(\beta h)}\right]^2\ket_{\bxi}
\end{eqnarray}
and similar expressions follow
for the observables $m^\mu_\tau$, $q_{\tau\tau}$ and $q_{\s\tau}$.

\section{The self-consistent equation and the free energy}
\label{sec:selfcons}

\subsection{Derivation of the self-consistent equation}

In order to arrive at a self-consistent equation for the densities
$W_{\bxi}(h,r,t)$ we first substitute the RS ansatz
(\ref{eq:RS}) into (\ref{eq:selfcons1}). Then, following \cite{bastian}, we isolate the
occurences of quantities of the form $\sum_{\alpha}\s_\alpha$,
$\sum_{\alpha}\tau_\alpha$ and
$\sum_{\alpha}\s_{\alpha}\tau_{\alpha}$ by inserting
\begin{eqnarray}
1
&=&
\sum_{m_\s=-\infty}^{\infty}\int_0^{2\pi}\frac{d \hat{m}_\s}{2\pi}\
e^{i\hat{m}_\s(m_{\s}-\sum_{\alpha}\s_\alpha)}
\\
1&=&\sum_{m_\tau=-\infty}^\infty\int_0^{2\pi}\frac{d \hat{m}_\tau}{2\pi}\
e^{i\hat{m}_\tau(m_{\tau}-\sum_{\alpha}\tau_\alpha)}
\\
1&=&\sum_{q=-\infty}^\infty\int_0^{2\pi}\frac{d\hat{q}}{2\pi}\
e^{i\hat{q}(q-\sum_{\alpha}\s_\alpha\tau_\alpha)}
\end{eqnarray}
After some algebra we can take the limit $n\to 0$ in our
equations and arrive at:
\begin{eqnarray}
\lefteqn{
\int dW_{\bxi}(h,r,t)\ e^{\beta hm_{\s}+\beta rm_\tau+\beta tq}}
\nonumber
\\
& &
\fl
=\exp\left[c\bra \int d W_{\bxi'}(h',r',t')\left(e^{m_\s K_1(h',r',t';\bxi\cdot\bxi')+m_\tau
K_2(h',r',t';\bxi\cdot\bxi')+q K_3(h',r',t';\bxi\cdot\bxi')}-1\right)\ket_{\bxi'}\right]
\label{eq:selfcons2}
\end{eqnarray}
where we have used the notation $dW_{\bxi}(h,r,t)=dhdrdt\ W_{\bxi}(h,r,t)$
and we introduced
\begin{equation}
K_1(h',r',t';\bxi\cdot\bxi')=\frac14\log\frac
{\Omega_{++}\Omega_{+-}}
{\Omega_{-+}\Omega_{--}}
\end{equation}
\begin{equation}
K_2(h',r',t';\bxi\cdot\bxi')=\frac14\log\frac
{\Omega_{++}\Omega_{-+}}
{\Omega_{+-}\Omega_{--}}
\end{equation}
\begin{equation}
K_3(h',r',t';\bxi\cdot\bxi')=\frac14\log\frac
{\Omega_{++}\Omega_{--}}
{\Omega_{+-}\Omega_{-+}}
\end{equation}
with
\begin{eqnarray}
\fl
\Omega_{\s\tau}\equiv
\Omega_{\s\tau}(h',r',t';\bxi\cdot\bxi')
&=&
4\cosh[\beta h'+\s\frac{\beta}{c}(\bxi\cdot\bxi')]\
\cosh[\beta r'+\tau\frac{\beta}{c}(\bxi\cdot\bxi')]\ \cosh[\beta
t']+
\nonumber
\\
\fl
&&
4\sinh[\beta h'+\s\frac{\beta}{c}(\bxi\cdot\bxi')]\
\sinh[\beta r'+\tau\frac{\beta}{c}(\bxi\cdot\bxi')]\ \sinh[\beta t']
\end{eqnarray}
Performing an inverse Fourier transform to (\ref{eq:selfcons2}),
expanding the right-hand
side and integrating over the magnetisations $m_\s,m_\tau$ and the
overlap $q$ we obtain
\[
\fl
W_{\bxi}(h,r,t)=\sum_{k=0}^\infty
\frac{e^{-c}c^k}{k!}
\left\bra\cdots\left\bra
\int[\prod_{l=1}^k dW_{\bxi^l}(h_l,r_l,t_l)]\
\delta\left[h-\frac{1}{\beta}\sum_{l=1}^k K_1(h'_l,r'_l,t'_l;\bxi\cdot\bxi^l)
\right]
\right.\right.
\]
\begin{equation}
\fl
\left.\left.\times\
\delta\left[r-
\frac{1}{\beta}\sum_{l=1}^kK_2(h'_l,r'_l,t'_l;\bxi\cdot\bxi^l)\right]
\delta\left[t-
\frac{1}{\beta}\sum_{l=1}^k K_3(h'_l,r'_l,t'_l;\bxi\cdot\bxi^l)\right]
\right\ket_{\bxi^1}\cdots\right\ket_{\bxi^k}
\label{eq:W}
\end{equation}
Expression (\ref{eq:W}) is the final result from which to evaluate
the densities $W_{\bxi}(h,r,t)$. Although appearing a daunting
numerical task it can be evaluated using the simple algorithm of
`population dynamics' as described e.g.\@ in \cite{mezardparisi}.
This relies on the local `tree-like' structure of
finitely-connected networks of spins and it consists of having a
large population of triplets $(h_l,r_l,t_l)$ which are to be
updated for a large number of iteration steps in the following way:
One first selects a number $k$ from a Poisson distribution of mean
$c$. Then, one chooses $l=1,\ldots,k$ triplets $(h_l,r_l,t_l)$ and
$l$ sublattices $\bxi^l$ at random and calculates the expressions
appearing in the delta functions. Finally, one selects a new
triplet $(h,r,t)$ and a new sublattice $\bxi$ at random and updates
them using the calculated expressions.

\begin{figure}[h]
\vspace{-15mm}
\setlength{\unitlength}{1.4mm}
\begin{picture}(120,55)
\put( 5,  0){\epsfysize=40\unitlength\epsfbox{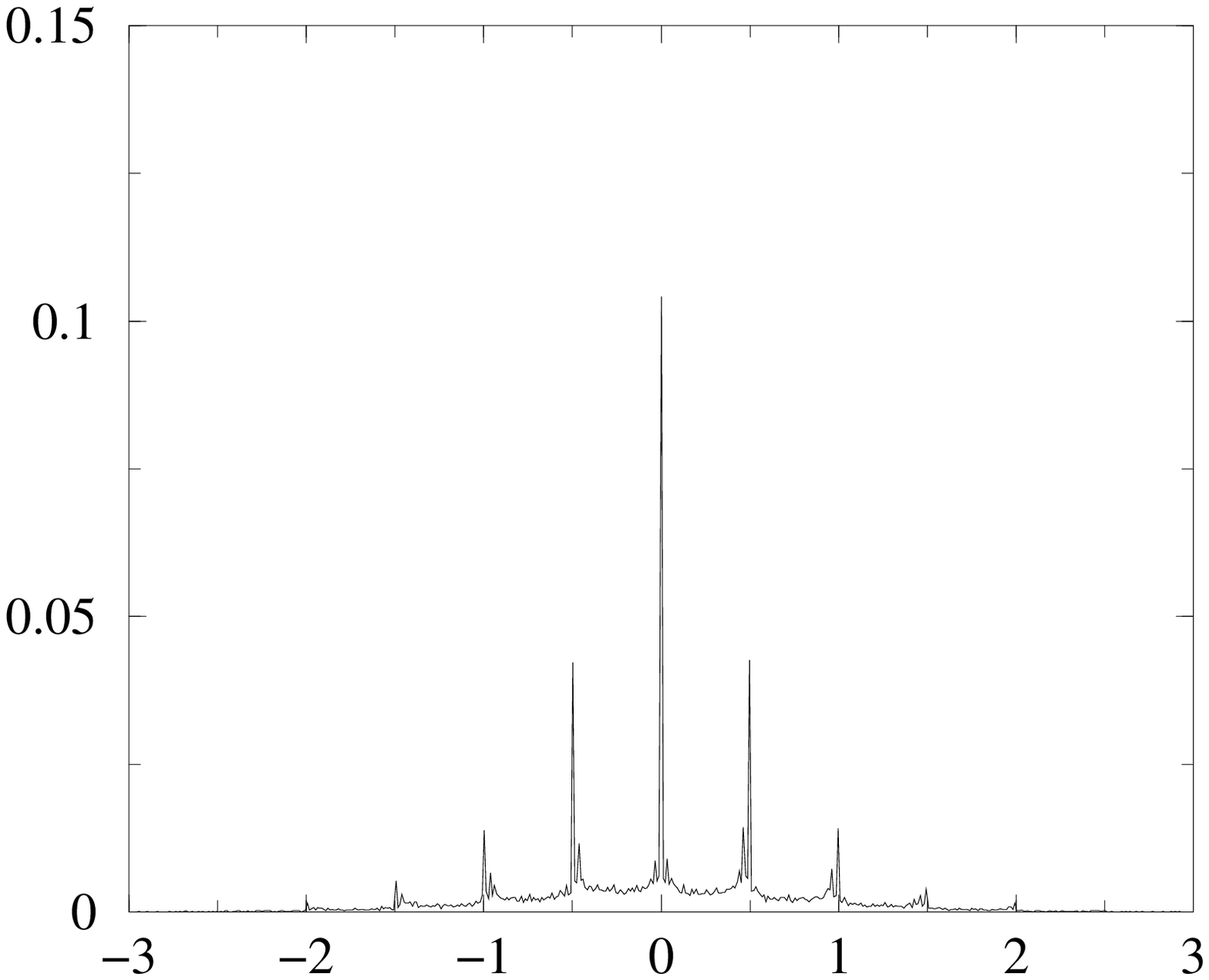}}
\put( 60,  0){\epsfysize=40\unitlength\epsfbox{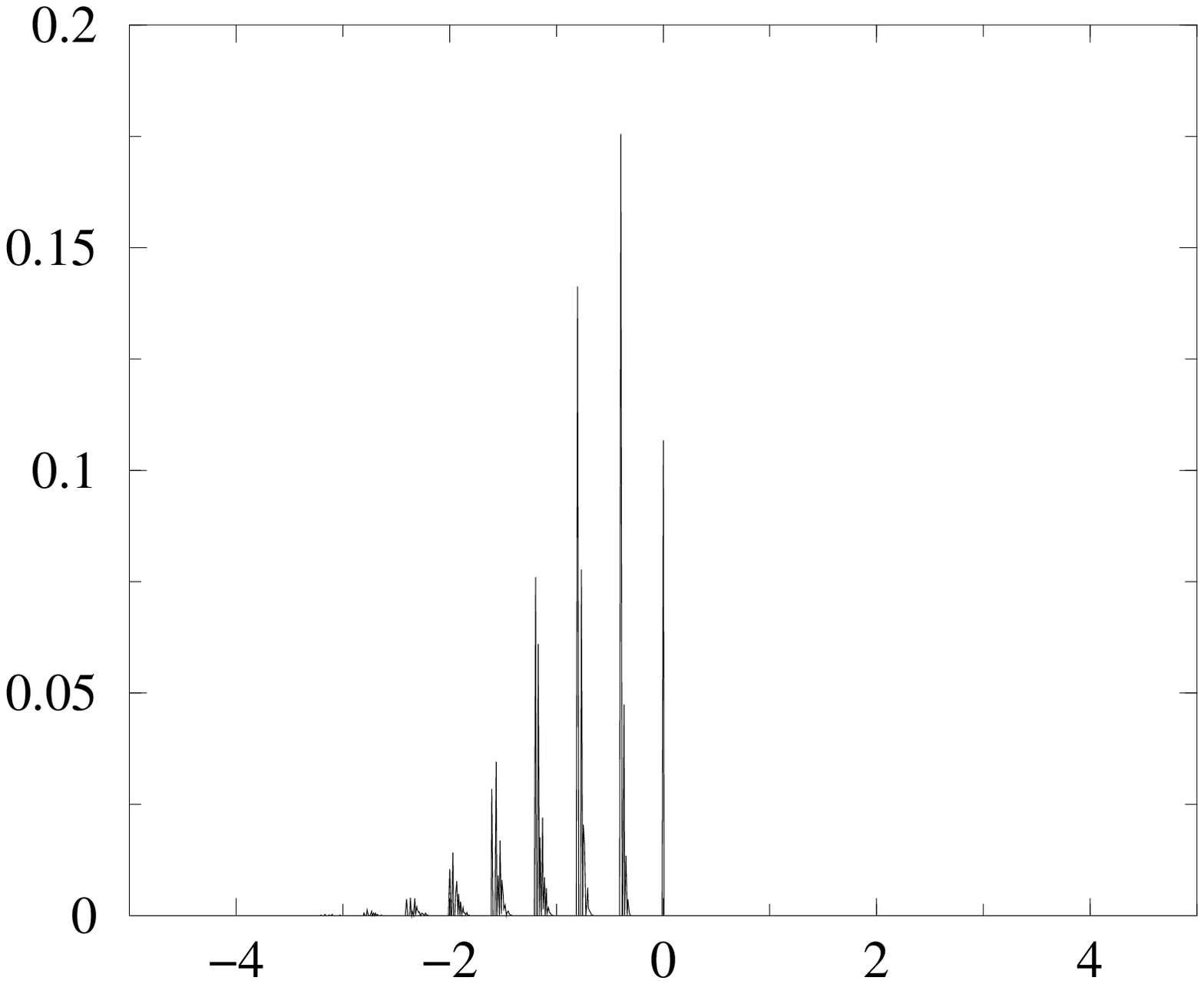}}
\put(0,20){$W^{h}_{\bxi}(h)$}
\put(31,-3){$h$}
\put(86,-3){$h$}
\end{picture}\vspace{5mm}
\caption{Typical profiles of the marginal densities $W^{h}_{\bxi}(h)$ for 
 $p=6, c=4, T=0.1$ (left picture, corresponding to the spin-glass region of figure \ref{fig:phase})
 and for $p=2, c=5, T=0.1$ (right picture, corresponding to the retrieval 
region of figure \ref{fig:phase}). 
Notice that since $p$ is even the peaks are located at $h=2l/c$ with $l$ integer.}
\label{fig:Wh}
\end{figure}

To get an idea about the profile of the densities $W_{\bxi}(h,r,t)$
we have applied the previous algorithm to study the marginal 
densities $W_{\bxi}^h(h)=\int drdt\,W_{\bxi}(h,r,t)$, 
$W_{\bxi}^r(r)=\int dhdt\, W_{\bxi}(h,r,t)$ and
$W_{\bxi}^t(t)=\int dhdr\, W_{\bxi}(h,r,t)$. Firstly, we observe
that close to the zero-temperature limit the marginals
$W_{\bxi}^h(h)$ and $W_{\bxi}^r(r)$ become a collection of delta
peaks; in this limit due to the absence of thermal noise the
effective fields $\{h,r\}$ become identical to the {\it true} local fields.
Since the average number of connections per neuron is here a finite number 
and the couplings
are discrete, the local fields are a multiple integer 
\cite{kanter}  of $1/c$ \footnote{
In fact, for even number of patterns $p$  we have that the local fields present peaks at $2l/c$
with $l$ integer, while for odd $p$ we have $l/c$.}
(see figure \ref{fig:Wh} for typical profiles of the marginal densities).  
As one moves away from the $T=0$ regime the
profile of the marginal densities $W_{\bxi}^h(h)$ and
$W_{\bxi}^r(r)$ takes a non-trivial form until the paramagnetic
phase is reached where one observes that
$W_{\bxi}^h(x)=W_{\bxi}^r(x)=W_{\bxi}^t(x)=\delta(x)$, as it
should. We have numerically evaluated the above marginals for several
regions of the parameter space and it is interesting to note that 
$W_{\bxi}^h(h)$ and $W_{\bxi}^r(r)$ are identical
(within the accuracy of numerical precision) 
\footnote{This is of course expected since
the hamiltonian (\ref{eq:f_def}) is invariant under the interchange of the two spin species.}
whereas $W_{\bxi}^t(t)=\delta(t)$.
The latter implies that our RS ansatz can be simplified
further. However this result relies on numerical observations and a
rigorous analytical proof appears somewhat hard
(one cannot rely on an induction proof since at the first iteration
step of (\ref{eq:W}) the fields $t$ must take a non-zero value). Nevertheless, 
it is interesting to note that setting $t=0$ to (\ref{eq:W}) 
completely decouples the density $W_{\bxi}(h,r,t)$ to 
two identical densities 
\[
\fl
Q_{\bxi}(x)=\sum_{k=0}^\infty
\frac{e^{-c} c^k}{k!}
\left\bra\left\bra\cdots\int[\prod_{l=1}^k dx_l\ Q_{\bxi}(x_l)]\
\right.\right.
\]
\begin{equation}
\left.\left.
\times\
\delta\left[x-\frac{1}{\beta}
\sum_{l=1}^k\atanh\left(\tanh[\beta x_l]\tanh[\frac{\beta}{c}
(\bxi\cdot\bxi^l)]\right)\right]\right\ket_{\bxi^1}\cdots\right\ket_{\bxi^k}
\end{equation}
with $x=\{h,r\}$. Note that this equation is now the same
as the one that follows from the analysis of \cite{bastian}, as it should.

\subsection{The free energy}
\label{sec:energy}

Let us now express the free energy (\ref{eq:f2}) as function of the
densities $W_{\bxi}(h,r,t)$. The first term comprising (\ref{eq:f2})
(energetic part) can be calculated without difficulty. One uses the
RS ansatz (\ref{eq:RS}) to replace the distributions
$P_{\bxi}(\bsigma,\btau)$ by the densities $W_{\bxi}(h,r,t)$ and
subsequently one performs the spins summations and takes the limit
$n\to 0$. The second term is slightly more complicated.
First we expand the exponential function. This replicates the
traces over the spins and sublattice variables. Inserting then for
each of the replicated densities the RS ansatz (\ref{eq:RS}) leads
to
\begin{eqnarray}
\fl
\lefteqn{\left\bra\log\sum_{\bsigma\btau}\exp\left[c\bra\sum_{\hat{\bsigma}\hat{\btau}}
P_{\bxi'}(\hat{\bsigma},\hat{\btau})
\left(e^{\frac{\beta J}{c}(\bxi\cdot\bxi')
\sum_\alpha(\s_\alpha\hat{\tau}_\alpha
+\hat{\s}_\alpha\tau_\alpha)}-1\right)\ket_{\bxi'}\right]\right\ket_{\bxi}}
\nonumber
\\
\fl
& &
=\left\bra\log\sum_{k=0}^\infty\frac{e^{-c}c^k}{k!}
\left\bra\cdots\left\bra\int[\prod_{l=1}^k
dW_{\bxi^l}(h_l,r_l,t_l)]
\right.\right.\right.
\\
& &
\left.\left.\left.
\times\
\prod_{\alpha=1}^n
\sum_{\s_\alpha\tau_\alpha}\prod_{l=1}^k\sum_{\hat{\s}_{\alpha}^l\hat{\tau}_\alpha^l}
e^{\frac{\beta}{c}(\bxi\cdot\bxi^l)(\s_\alpha\hat{\tau}_\alpha^l+\hat{\s}_\alpha^l\tau)+\beta
h_l\hat{\s}_\alpha^l+\beta\hat{\tau}_\alpha^l+\beta
t_l\hat{\s}_\alpha^l\hat{\tau}_\alpha^l}
\right\ket_{\bxi^1}\cdots\right\ket_{\bxi^k}\right\ket_{\bxi}
\nonumber
\end{eqnarray}
Performing the spin summations in the above equation and
using the simple expression $F(\s_\alpha,\tau_\alpha)=
\sum_{\s\tau}\delta_{\s,\s_\alpha}\delta_{\tau,\tau_\alpha}
F(\s,\tau)$ to relocate the occurences of $\s_\alpha$ and $\tau_\alpha$,
we can take the limit $n\to 0$. The final result for the free
energy (\ref{eq:f2}) is then
\[
\fl
f=
-\frac{1}{\beta}\sum_{k=0}^\infty\frac{e^{-c}c^k}{k!}
\left\bra\left\bra\cdots\left\bra\int[\prod_{l=1}^k dW_{\bxi^l}(h_l,r_l,t_l)]\
\log \left[\sum_{\s\tau=\pm}\prod_{l=1}^k
\frac{\Omega_{\s\tau}(h_l,r_l,t_l;\bxi\cdot\bxi^l)}
{\mathcal{N}(h_l,r_l,t_l)}\right]
\right\ket_{\bxi^1}\cdots\right\ket_{\bxi^k}\right\ket_{\bxi}
\]
\[
\fl
+\frac{c}{2\beta}\left\bra \int dW_{\bxi}(h,r,t)dW_{\bxi'}(h',r',t')\right.
\]
\begin{equation}
\fl
\left.
\log\left[\frac{\sum_{\tau\tau'=\pm}
\cosh[\beta h+\beta t\tau+\frac{\beta}{c}(\bxi\cdot\bxi')\tau']
\cosh[\beta h'+\beta t'\tau'+\frac{\beta}{c}(\bxi\cdot\bxi')\tau]
e^{\beta r\tau+\beta r'\tau'}}
{\sum_{\tau\tau'=\pm}
\cosh[\beta h+\beta t\tau]
\cosh[\beta h'+\beta t'\tau]
e^{\beta r\tau+\beta r'\tau'}}
\right]\right\ket_{\bxi\bxi'}
\label{eq:f3}
\end{equation}

\section{Phase Diagrams}
\label{sec:results}

\subsection{Bifurcation Analysis}

The numerical observations concerning the 
densities $W_{\bxi}(h,r,t)$ in section \ref{sec:selfcons}
support the equivalence between the parallel and 
sequential versions of the finite-$c$ Hopfield model. 
In this section we will show further that one can analytically prove 
that the location of the second-order transitions between 
paramagnetic/spin-glass
and paramagnetic/retrieval phases is identical to those of sequential 
dynamics.

Our bifurcation analysis is similar in spirit to \cite{bastian}.
First we note that $W_{\bxi}(h,r,t)=\delta(h)\delta(r)\delta(t)$
solves (\ref{eq:W}) for all $\bxi$. This solution, as can be easily confirmed,
corresponds to the high-temperature paramagnetic state where no recall is 
possible and $m_{\bxi}^\s=m_{\bxi}^\tau=0$.
As the temperature is
lowered from $T=\infty$ we expect non-trivial solutions to appear.
To determine the transition temperature at which these occur we
first assume that close to the transition $\int dW_{\bxi}(h,r,t)h^\ell
=\mathcal{O}(\epsilon^\ell)$, $\int dW_{\bxi}(h,r,t)r^\ell
=\mathcal{O}(\epsilon^\ell)$ and $\int dW_{\bxi}(h,r,t)t^\ell
=\mathcal{O}(\epsilon^\ell)$. We can then expand equation
(\ref{eq:selfcons2}) and identify term by term each expression in the
resulting power series in the left- and right-hand
sides of (\ref{eq:selfcons2}). To second order, this one-to-one
correspondence firstly indicates that integrals carrying the field $t$ 
must vanish. We then obtain:
\begin{equation}
\fl
\int dW_{\bxi}(h,r,t)\ h=c\bra \int dW_{\bxi'}(h,r,t)\ h\
\tanh[\frac{\beta}{c}(\bxi\cdot\bxi')]\ket_{\bxi'}
\end{equation}
\begin{equation}
\fl
\int dW_{\bxi}(h,r,t)\ r=c\bra \int dW_{\bxi'}(h,r,t)\ r\
\tanh[\frac{\beta}{c}(\bxi\cdot\bxi')]\ket_{\bxi'}
\end{equation}
\begin{equation}
\fl
\int dW_{\bxi}(h,r,t)\, h^2-\left[\int dW_{\bxi}(h,r,t)\, h\right]^2
=c\bra \int dW_{\bxi'}(h,r,t)\,
h^2\,\tanh[\frac{\beta}{c}(\bxi\cdot\bxi')]\ket_{\bxi'}
\end{equation}
\begin{equation}
\fl
\int dW_{\bxi}(h,r,t)\, r^2-\left[\int dW_{\bxi}(h,r,t)\, r\right]^2
=c\bra \int dW_{\bxi'}(h,r,t)\,
r^2\,\tanh[\frac{\beta}{c}(\bxi\cdot\bxi')]\ket_{\bxi'}
\end{equation}
These four equations mark the different types of transitions
away from the paramagnetic state. The first pair of equations
will give us the transition from $m_{\bxi}^\s=0$ and $m_{\bxi}^\tau=0$
respectively to a non-trivial recall (ferromagnetic) state
whereas the second pair will give us the transition to the spin-glass
state. The resulting similarity in the first and second pair
of the above equations is a consequence of the equivalence of the
two species of spins in our system.

To determine now the critical
parameter values for which these transitions occur we need to
evaluate the highest temperature for which the following
$2^p\times 2^p$ matrices have an eigenvalue equal to 1:
\begin{equation}
M_{\bxi\bxi'}=c p_{\bxi'}\tanh[\frac{\beta}{c}(\bxi\cdot\bxi')]
\hspace{10mm}
Q_{\bxi\bxi'}=c p_{\bxi'}\tanh^2[\frac{\beta}{c}(\bxi\cdot\bxi')]
\end{equation}
These equations are identical to those found in \cite{bastian}.
An elegant construction of the eigenvalues of
matrices of the above form (which relies on exploiting the
properties $M_{\bxi\bxi'}=M(\bxi\cdot\bxi')$ and
$Q_{\bxi\bxi'}=Q(\bxi\cdot\bxi')$) has already been given
in \cite{vanhemmen}. Here we state the final result determining
the transition lines, identical to that found in \cite{bastian}:
\begin{eqnarray}
{\rm P\to R:} & &
\hspace{4mm}
\frac{c}{2^p\, p}\sum_{m=0}^p
\left(\!\!\begin{array}{c}
p\\ m
\end{array}\!\!\right)
(p-2m)\tanh[\frac{\beta}{c}(p-2m)]=1
\label{eq:PF}
\\
{\rm P\to SG:} & &
\hspace{4mm}
\frac{c}{2^p}\sum_{m=0}^p
\left(\!\!\begin{array}{c}
p\\ m
\end{array}\!\!\right)
\tanh^2[\frac{\beta}{c}(p-2m)]=1
\label{eq:PSG}
\end{eqnarray}

In general, the bifurcation analysis can not be applied 
to determine the spin-glass/retrieval transition line 
which must be computed directly from equations (\ref{eq:W}) and (\ref{eq:ms}). 
This numerical task becomes increasingly difficult for large values of the 
connectivity parameter $c$ (since one is required to evaluate $2^p$ 
sublattice densities
where $p\simeq c$ close to the spin-glass/retrieval transition).
For sufficiently small values of the connectivity parameter $c$ (up to $c=7$) 
we find that this transition line is identical to that of sequential dynamics. 
It is interesting to note that within replica-symmetric considerations, 
the $c\to\infty$ (extremely-diluted) phase 
diagrams of sequential and parallel Hopfield models are not identical, 
indicating that
a critical $c_\star$ exists above which $W_{\bxi}^t(t)$ 
is no longer given by $\delta(t)$.
Determining this transition is a challenging numerical task.

\subsection{Results and comparison with simulations}

\begin{figure}[t]
\vspace{-15mm}
\setlength{\unitlength}{1.4mm}
\begin{picture}(120,55)
\put( 5, 0){\epsfysize=40\unitlength\epsfbox{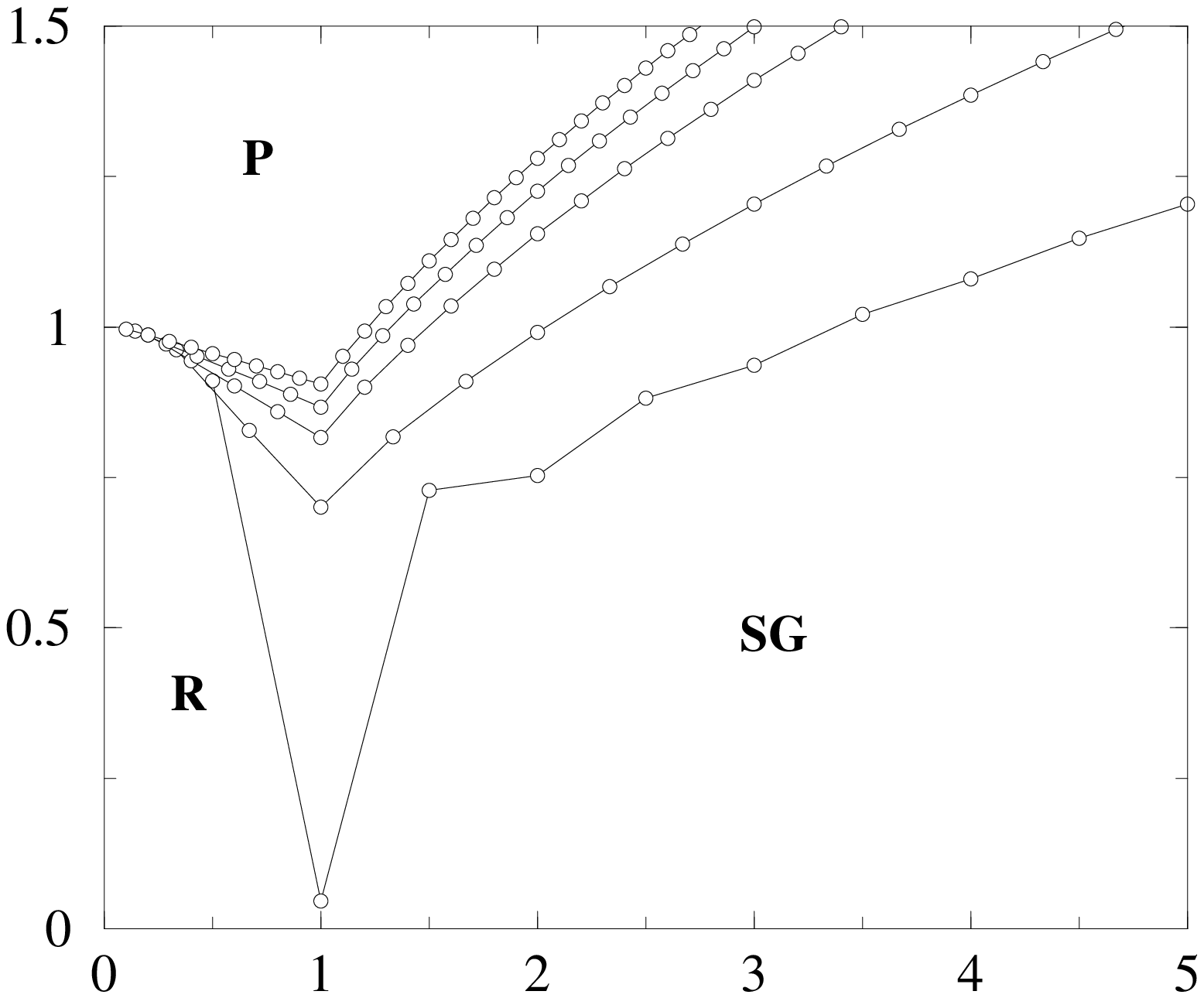}}
\put( 60, 0){\epsfysize=40\unitlength\epsfbox{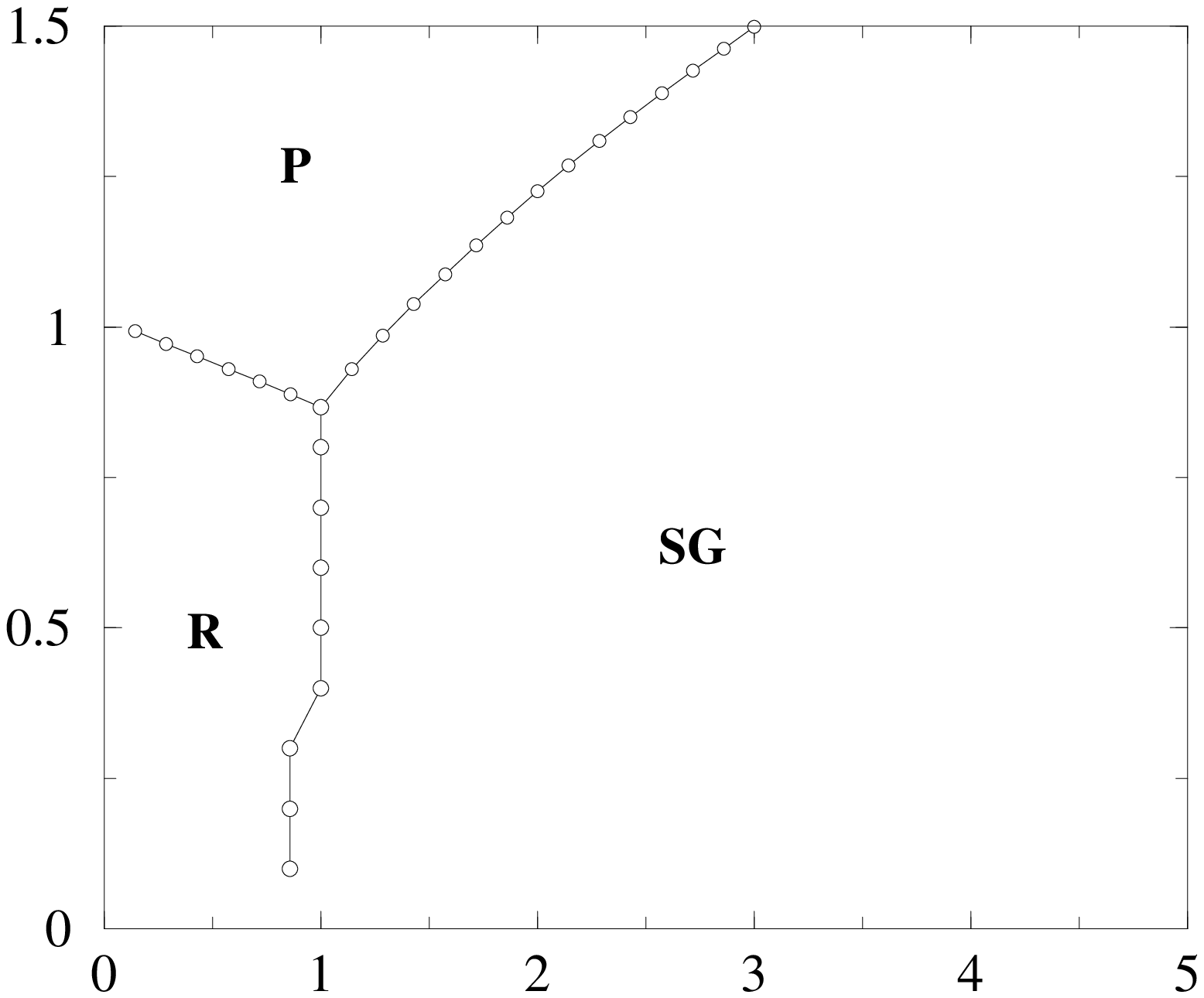}}
\put(0,20){$T$}
\put(31,-3){$\alpha=p/c$}
\put(86,-3){$\alpha=p/c$}
\end{picture}\vspace{5mm}
\caption{Phase diagram of the finite-connectivity Hopfield model
with synchronous dynamics in the $(\alpha,T)$ plane. Left: The
paramagnetic/retrieval and paramagnetic/spin-glass 
transition lines as obtained from the bifurcation analysis.
(from lower to upper lines $c=\{2,3,5,7,10\}$).
Right: The $c=7$ phase diagram with the spin-glass/retrieval transition line (obtained
by direct evaluation of the pattern overlaps (\ref{eq:ms})
using population dynamics).
Markers correspond to integer $p$ values (lines are simply put as guides to 
the eye). \emph{On} the retrieval/spin-glass transition line $m_\s=m_\tau=0$.}
\label{fig:phase}
\end{figure}

By numerically solving equations (\ref{eq:ms},\ref{eq:W})
we find that only one  magnetisation component (also 
called pattern overlap) $m^\mu_\lambda$ (\ref{eq:mlambda})
provides non-trivial solutions (i.e.\@ only one pattern 
is retrieved while the others are zero.). For notational simplicity, 
we will from now on denote these non-zero  solutions simply as 
$m_\s$ and $m_\tau$.

Plotting solutions of (\ref{eq:PF}) and (\ref{eq:PSG}) in the
$(\alpha,T)$ plane (with $\alpha=p/c$) results in the phase diagram
of figure \ref{fig:phase}. It has been plotted for
$c\in\{2,3,5,7,10\}$ and it consists of three phases: a high-temperature 
paramagnetic phase with all observables identically zero, 
a retrieval phase with $m_\s,m_\tau\neq 0$ and
$q_{\lambda\lambda'}=0$ (with $\lambda=\s,\tau$) 
and a spin-glass phase with $m_\s=m_\tau=0$ 
and $q_{\lambda\lambda'}\neq 0$ 
\footnote{note that 
due to the inequality $q_{\alpha_1\alpha_2}\geq q_{\alpha_1
\alpha_2\alpha_3\alpha_4}
\geq \cdots \geq 0$ we only need to examine $q_{\alpha_1\alpha_2}$ 
to identify the spin-glass phase.}. For $T\leq 1$ 
the spin-glass/retrieval transition is obtained numerically by
direct evaluation of our observables.

We see that already for $c=10$
(upper line) the phase diagram resembles closely the one derived in
the extremely-diluted model \cite{canning} where the transition
lines are given by $T=1$ for $\alpha\leq 1$ (P$\to$R) and
$T=\sqrt{\alpha}$ for $\alpha\geq 1$ (P$\to$SG). From the
perspective of neural network operation, this shows that sparsely
connected models retain the same qualitative features for a wide
variety of synaptic connections per neuron and can retrieve
information even for surprisingly small values of connections.

\begin{figure}[t]
\vspace{-15mm}
\setlength{\unitlength}{1.4mm}
\begin{picture}(120,55)
\put( 30,  0){\epsfysize=45\unitlength\epsfbox{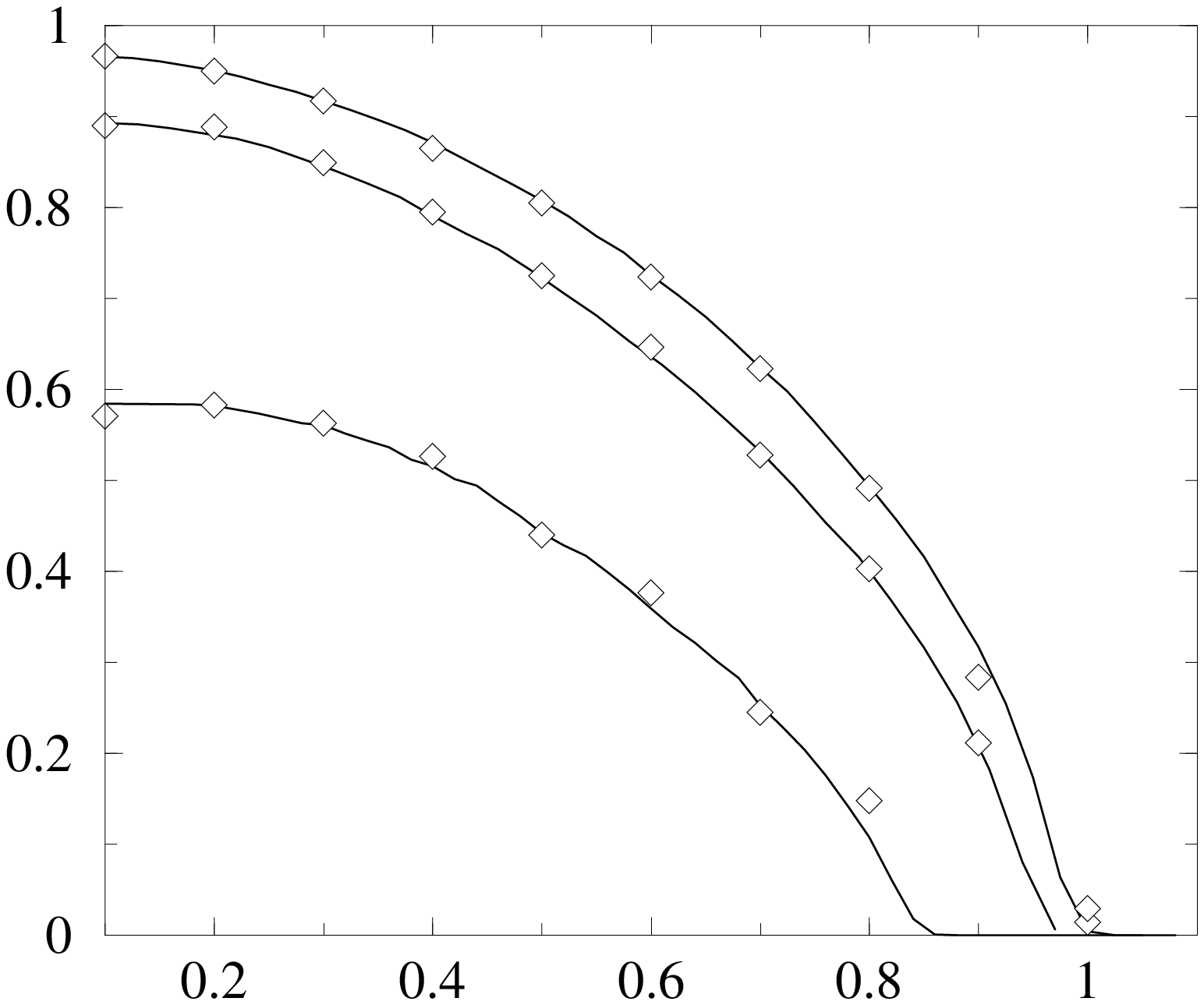}}
\put(25,25){$m$}
\put(58,-3 ){$T$}
\end{picture}
\caption{Pattern overlaps $m_\s=m_\tau=m$
as functions of $T=1/\beta$
calculated from equations (\ref{eq:ms}) and (\ref{eq:W}).
Parameter values are: $p=2$ and $c=\{3,5,7\}$ (from lower to upper lines).
Markers correspond to simulation experiments of $N=10,000$ neurons.}
\label{fig:magn}
\end{figure}

In figure \ref{fig:magn} we present solutions of the sublattice
overlap $m_\s$ (\ref{eq:ms}) and $m_\tau$ showing
the second-order transition from retrieval to paramagnetic states
for the simplest non-trivial case of $p=2$. These have been drawn
for $c=\{3,5,7\}$ showing the effect of different degrees of
connectivity on the retrieval success. 
To verify our results we have performed simulation experiments on
the dynamical process (\ref{eq:micro_dynamics}) for a system size
of $N=10, 000$ neurons (markers of figure \ref{fig:magn}). These
appear to be in very good aggrement with our theory. The simulation
experiments also show that while Hebbian-type couplings
(\ref{eq:interactions}) lead to fixed-point stationary solutions,
anti-Hebbian ones where
$J_{ij}=-\frac{c_{ij}}{c}(\bxi_i\cdot\bxi_j)$ lead to 2-cycles with
$m_\s(t)=-m_\s(t+1)$ (and similarly for $m_\tau$).
Also, evaluation of the free energy (\ref{eq:f3}) as a 
function of temperature $T$ 
shows that it is a monotonically decreasing function
indicating that the entropy remains positive even for 
low temperatures. 

We have also compared our pattern overlaps with those derived
from the analysis of \cite{bastian} (which has as a starting the
Ising `sequential' Hamiltonian) and within  the accuracy of numerical precision 
we have found that observables
of the two systems are identical. This result, although somewhat
expected given both knowledge from earlier neural network studies
and, of course, the identity of our bifuraction results with those
of \cite{bastian}, is also a bit surprising: in the process of
solving our equations we introduced a 3D RS `effective-field'
distribution (compared to the 1D sequential one of \cite{bastian}).
As it turns out however the system in
the process of updating fields by iteration of (\ref{eq:W})
effectively discards those which describe species-correlations and
factorises $W_{\bxi}(h,r,t)$  to $W_{\bxi}(h,r)\,\delta(t)$.
Henceforth, our 3D RS field-distribution  reduces to a 2D one. The
equivalent treatment of $\bsigma$ and $\btau$ then ensures (as also
in analytically simpler models) that equilibrium observables
between sequential and synchronous systems are identical (this relation 
however ceases to exist for a certain 
large value $c_\star$ since for $c\to\infty$ 
the sequential and parallel phase diagrams \emph{are} different \cite{toni}).
\begin{figure}[h]
\vspace{-15mm}
\setlength{\unitlength}{1.4mm}
\begin{picture}(120,55)
\put( 30,  0){\epsfysize=45\unitlength\epsfbox{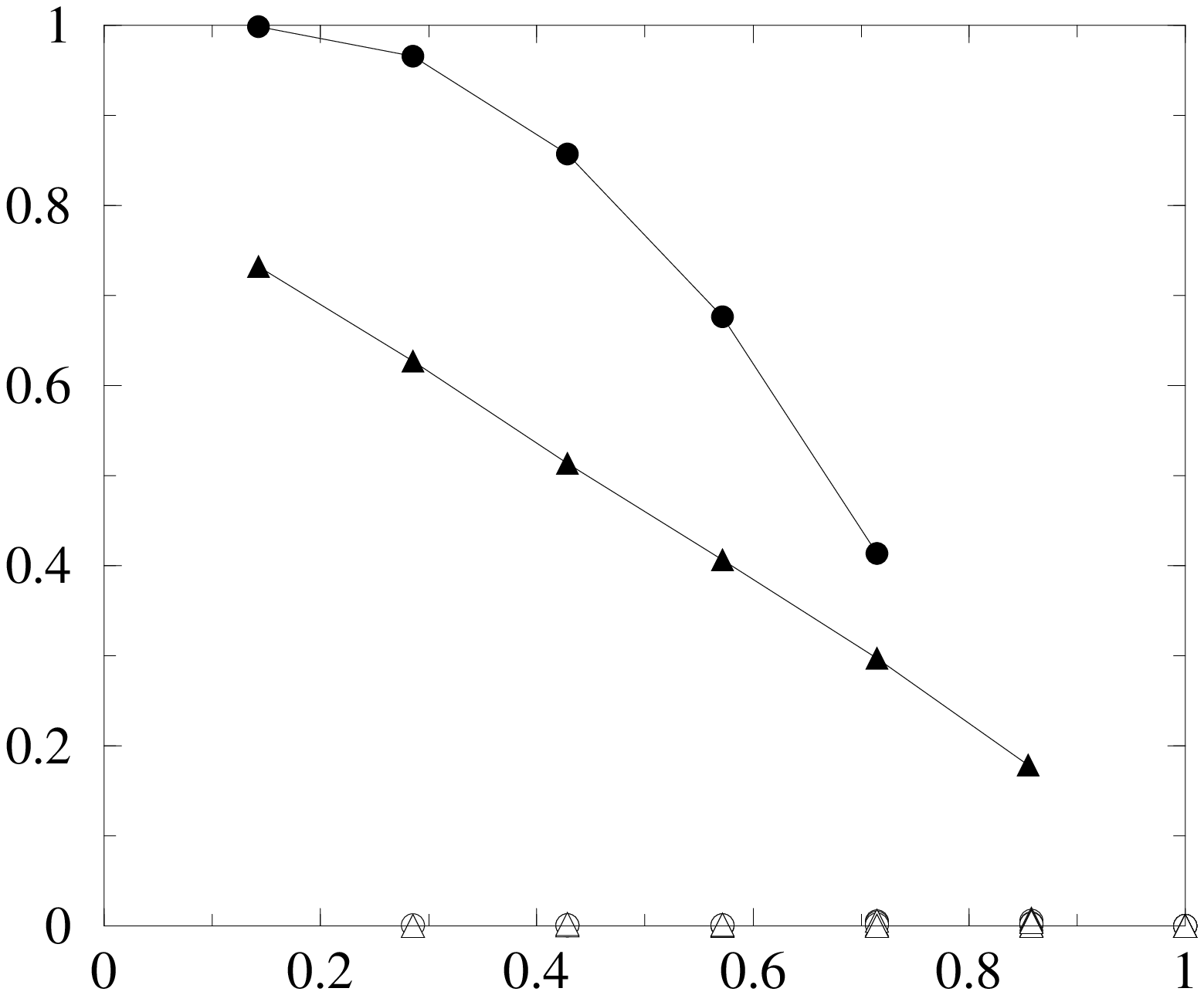}}
\put(25,25){$m^\mu$}
\put(58,-2 ){$\alpha$}
\end{picture}
\caption{Pattern overlap $m^\mu_\s=m^\mu_\tau=m^\mu$ 
as a function of $\alpha$ (with $c=7$) for $T=0.1$ 
(circles) and $T=0.7$ (triangles). 
Here we can see the first-order transition from retrieval 
to the spin-glass phase at $\alpha=6/7$ and $\alpha=7/7$ for
$T=0.1$ and $T=0.7$ respectively. }
\label{fig:sgr}
\end{figure}

In figure \ref{fig:sgr} we have plotted the pattern overlaps $m^\mu_\s$
and $m^\mu_\tau$ as
a function of $\alpha$ for different values of the temperature
(note that $m_\s^\mu=m_\tau^\mu$) . We have set
the average connectivity to $c=7$ while varying the number of patterns $p$ 
(therefore the number of pattern overlaps $m^\mu$ also changes with $\alpha$). 
From this graph we can complete the phase diagram (figure \ref{fig:phase}), 
determining the location of the first-order transition from retrieval to spin-glass
states.

\section{Discussion}

In this paper we have studied equilibrium properties of 
attractor neural network models with finite connectivity 
in which neurons operate in a parallel way.
This work is motivated by the interesting properties that the two types
of dynamical models share in simpler (e.g.\@ fully-connected) scenarios: 
There, and for a surprisingly large number of universality classes,
one can prove analytically that the equilibrium states 
following from the two different dynamical rules are identical. 
Our starting point here is the Peretto Hamiltonian (\ref{eq:H}). 
We have followed on the footsteps of \cite{bastian} to derive
the transition lines in our phase diagram
and expressions for the (sublattice) overlaps.
The resulting phase diagram plotted in the $(\alpha,T)$ plane 
then consists of three phases: a high-temperature 
paramagnetic phase, a retrieval and a spin-glass phase.
The transitions from the paramagnetic phase have been determined 
analytically while the first-order retrieval/spin-glass transition has been computed 
using population dynamics \cite{mezardparisi}.
Under our replica-symmetric considerations we have shown that 
the retrieval properties of the parallel
finite-connectivity Hopfield model are identical to those
of the sequential one. Comparison with numerical simulations for large system 
sizes shows excellent agreement. 

Many questions remain to be answered for neural
networks of finite connectivity. Using the same framework 
(as initiated by \cite{bastian}) 
one can proceed further with the study of multi-state
e.g.\@ $Q$-Ising or Blume-Emery-Griffiths neural networks.
A different appoach would be to study systems which consist 
of two species of operating units (as, effectively, here) 
but which explicitly interact with one another. 
This would lead us to an Ashkin-Teller-type
neural network in which the phase diagram can be different.
Non-trivial extensions would be to study the validity of the RS solution following
the schemes in \cite{monasson,montanari}. These will be the subject of a future work.

\section*{Acknowledgments}

We are indebted to Bastian Wemmenhove and Toni Verbeiren for 
very helpful discussions and the Fund for Scientific
Research-Flanders, Belgium. IPC thanks The Abdus Salam ICTP
for hospitally and NSS wishes to thank 
Professor D. Boll\'e for supporting this research.

\end{document}